\documentclass[prd,twocolumn,tightenlines,superscriptaddress,preprintnumbers,floatfix,nofootinbib,]{revtex4}

\usepackage{graphicx}
\usepackage{amsmath}
\usepackage{amssymb}
\usepackage{bbold}
\usepackage{color}
\input epsf

\newcommand{\qhat}{\hat{q}} 

\begin{document}

\title{Picturing perturbative parton cascades in QCD matter}
\author{Aleksi Kurkela and Urs Achim Wiedemann}
\preprint{CERN-PH-TH-2014-121}
\affiliation{Physics Department, Theory Unit, CERN, CH-1211 Gen\`eve 23, Switzerland}
\begin{abstract}
Based on parametric reasoning, we provide a simple dynamical picture of how a perturbative parton cascade, in interaction with a QCD medium, fills phase space as a function of time. 
\end{abstract}

\maketitle 

\section{Introduction}
There are essentially two motivations for studying how a perturbative parton shower, embedded in QCD 
matter, evolves on all energy scales and on all angular scales. First, this process has been 
recognized since long as a useful set-up for elaborating the 
dynamics of thermalization in QCD~\cite{Baier:2000sb}.
This is so, since the parton distribution characterizing a jet is
initially far from equilibrium, but in a thermal QCD medium it will evolve at late times into a distribution 
that is indistinguishable from a thermal one. In this context, recent detailed analyses have determined
for instance that the thermalization time obtained from a weak coupling 
treatment~\cite{Kurkela:2011ti,Kurkela:2011ub,Kurkela:2014tea} is comparable to that established
for non-abelian gauge theories in the strong coupling limit~\cite{Gubser:2008as,Hatta:2008tx,Chesler:2008uy,Chesler:2010bi}.
Secondly, studying the medium-modifications of perturbative parton showers is motivated, of course, by
the recent measurements of reconstructed jets in heavy ion collisions at the LHC~\cite{Chatrchyan:2011sx,Chatrchyan:2012gw,Chatrchyan:2013kwa,Chatrchyan:2014ava,Aad:2010bu,Aad:2012vca,Aad:2014wha,Abelev:2013kqa} that characterize in unprecedented detail the medium-induced redistribution of jet energy and jet quanta in longitudinal (\emph{i.e.} along the jet axis) and transverse (\emph{i.e.} orthogonal to the jet axis) phase space. While it is conceivable that strong coupling techniques are needed to explain these data~\cite{Casalderrey-Solana:2014bpa}, the observation of modified but vacuum-like fragmentation
patterns~\cite{Chatrchyan:2013kwa,Chatrchyan:2014ava} in high energy jets at the LHC gives support to approaches that formulate medium effects within a perturbative framework~\cite{Zapp:2012ak}. Recent approaches extend
the early perturbative formulation of medium-induced parton energy loss~\cite{BDMPSZ,Zakharov:1996fv,Wiedemann:2000za,Gyulassy:2000er,Wang:2001ifa,Arnold:2002ja,Arnold:2009ik} in particular by better 
analyzing the role of color coherence and transverse broadening in the evolution of the medium-modified
perturbative parton cascade~\cite{CaronHuot:2008ni,CasalderreySolana:2011rz,MehtarTani:2011gf,Beraudo:2012bq,CasalderreySolana:2012ef,Blaizot:2013hx,Blaizot:2013vha,Panero:2013pla,D'Onofrio:2014qxa}, and by developing full 
Monte Carlo models for medium-modified parton showers~\cite{Schenke:2009gb,Armesto:2009fj,Zapp:2013vla}. 

%%%%%%%%%%%%%%%%%%%%%%%%%%%%%%%%%%%%%%%%%%
\begin{figure}[h!]
\begin{center}
\includegraphics[width=0.45\textwidth]{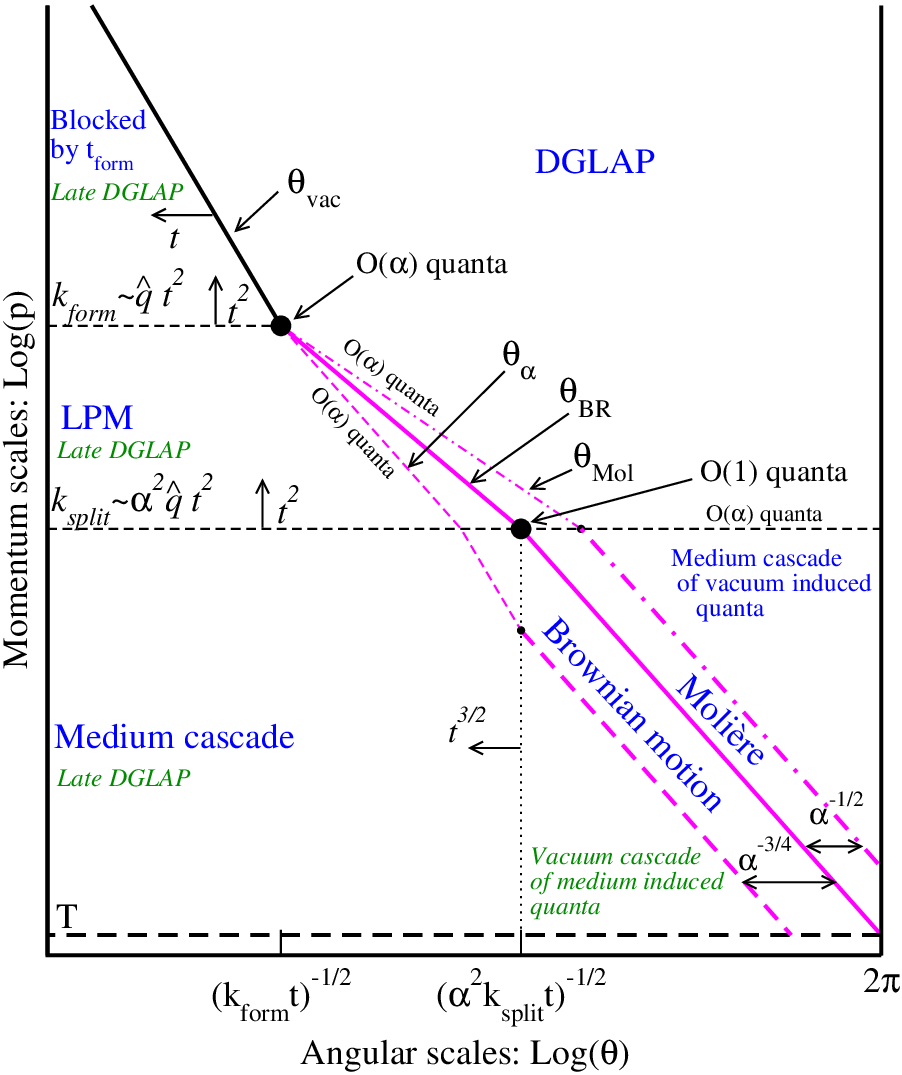}
\caption{
Parametrically accurate picture of how a medium-modified parton cascade fills
phase space. At time $t$, quanta can be formed up to momentum scale $k_{\rm form}$
and they are formed with $O(1)$ probability per $\log p$ at lower scale $k_{\rm split}$.
Quanta below $k_{\rm split}$ split further and their energy cascades to the thermal
scale $T$ in less than an epoch $t$. Transverse Brownian motion moves quanta
up to the angle $\theta_{\rm BR}(p)$ denoted by the thick purple line. 
The Moli\`ere region at larger $\theta$ is dominated by rare large angle scattering. At even
larger angle, there are $O(\alpha_s)$ quanta per double logarithmic phase space 
from DGLAP 'vacuum' radiation, and for momenta below $k_{\rm split}$ these
cascade within time $t$ to $T$. After the jet escapes the medium, the jet and
the emitted fragments will undergo vacuum radiation. This late time vacuum
radiation emitted by the original parton dominates at sufficiently small $\log \theta$  (regions marked ``late DGLAP'' and bounded by $\theta_{\rm vac}$ and $\theta_\alpha$), 
whereas the late time radiation of the fragments dominates in the region 
denoted by ``Vacuum cascade of the medium induced quanta''. Details given in the text. 
\label{fig1}
}
\end{center}
\end{figure}
%%%%%%%%%%%%%%%%%%%%%%%%%%%

Remarkably, the recent developments towards a jet quenching phenomenology applicable to
all jet energy and angular scales,
as well as the recent studies of the conceptually related thermalization problem, have identified 
largely independent of each other a set of parametric momentum scales and angular scales 
that characterize different aspects of the jet quenching phenomenon.
The present work grew out of our attempt to combine, what is parametrically known from these
studies, into a simple dynamical picture of how a perturbative parton cascade embedded in thermal QCD
matter fills phase space as a function of time. In elaborating this picture, we came across
parametric estimates of physics effects that we had not seen discussed before, such as 
parametric estimates for the interplay between vacuum and medium-induced radiation, for 
the medium-induced cascading of DGLAP vacuum radiation to lower momentum scales, 
for the late time (\emph{i.e.} after the jet has left the medium) vacuum cascading of medium-induced radiation, and
for the angular scales at which rare large angle scattering dominates over multiple soft
angle scattering. Including these effects, we provide in section~\ref{sec2} a simple,
kinematically complete and parametrically correct picture of the evolution of a perturbative 
parton cascade in QCD matter. We then illustrate the use of this
picture by relating parametric estimates of the jet energy fraction of different kinematical
regions to characteristic features in the measurements of quenched jets.  

\section{Jet evolution in the $\log p$-$\log \theta$-plane}
\label{sec2}
In this section, we motivate Fig.~\ref{fig1} that illustrates what happens parametrically
in perturbation theory when a jet propagates through a thermal cloud of temperature $T$. 
We consider a set up in which, at very early time $t=1/Q \sim 0$, a hard parton distribution 
$\sim \delta(\theta)\, \delta(p-Q)$ is embedded in the thermal bath. This `jet' 
is localized in angle $\theta$, and it lives at time $t = 0$ on a momentum scale $Q$ that
is much larger than any other scale in the problem. Subsequently, the jet evolves via 
perturbative parton branching. Fig.~\ref{fig1} depicts how its fragments (`splittees') fill 
the logarithmic phase space in momentum $p$ and angle $\theta$ as a function of time. 
We aim at a discussion that is based on minimal assumptions about the nature of the 
QCD matter. For simplicity, the medium is assumed to be time-independent, and - for most
of our arguments - it is assumed to be characterized solely by the transport coefficient
$\hat{q}$ that denotes the medium-induced squared transverse momentum broadening of 
energetic partons per unit path-length. Whenever our estimates rely on more detailed information
about the microscopic structure of the QCD medium, we shall state this explicitly.

\subsection{DGLAP region}
Quantum mechanical formation time prevents emission faster
than a time it takes to separate the wave packets of the splitter and the splittee.
In general, splittees at a scale $p<Q$ can form at a time $t\geq p/k_\perp^2$ set by their inverse transverse
energy.   For jet evolution in the vacuum, the angular distribution is determined by the 
primary decay kinematics of the splittee, $t \geq p/k_\perp^2 \sim 1/(p\, \theta^2)$.
Therefore, the logarithmically ordered DGLAP (Dokshitzer-Gribov-Lipatov-Altarelli-Parisi) `vacuum' parton shower
fills the $\log p$-$\log \theta$-plane from the outside in. In the entire region determined 
by 
$\theta > Ê\theta_{\rm vac}$ with
\begin{equation}
\theta_{\rm vac} \sim  1/(p t)^{1/2},
\end{equation}
 one finds with a probability $O(\alpha_s)$ per logarithmic phase space quanta due
to vacuum radiation,
\begin{equation}
\frac{d P_{\rm find}}{d \log\, p \, d \log \, \theta} \sim \alpha_s \, .
\label{eq1}
\end{equation}
The region where this primary splitting gives the dominant contribution 
in the $\log p$-$\log \theta$-plane is marked as the DGLAP-region in Fig.~\ref{fig1}.

\subsection{LPM region}
In contrast, medium-induced parton branching fills the 
$\log p$-$\log \theta$-plane from the bottom up (in $p$) and from the inside out (in $\theta$)~\cite{BDMPSZ}. 
This is so since transverse momentum 
is acquired by Brownian motion in the medium, $k_\perp^2 \sim \hat{q}\, t$; the formation
time constraint $t \geq p/k_\perp^2 \sim p/(\hat{q}\, t)$ implies then that 
medium-induced quanta can be formed in the region $p \lesssim k_{\rm form}$ where
\begin{equation}
 k_{\rm form}(t) \equiv \hat{q}\, t^2\, ,
\end{equation}
or, alternatively, that quanta at a scale $p$ can be formed at times $t > t_{\rm form}(p)$
where
\begin{equation}
 t_{\rm form}(p) \equiv \sqrt{\frac{p}{\hat{q}}}\, .
 \end{equation}
These quanta are created at small angles $\theta \lesssim \sqrt{\alpha_s} $, and
to our purposes we can treat the emitted quanta as being collinear with respect to the emitter.%
\footnote{The angle at which a quantum is created is $\theta^2\sim \hat{q} t_{\rm form}/p^2$, which for
a perturbative medium $\hat{q}\sim \alpha_s^2 T^3$ reads $\theta^2 \sim \alpha_s (T/p)^{3/2}$.}
Their angular distribution will be determined by reinteractions with the medium, discussed in section \ref{sec:F}.

A quantum can be formed at the scale $k_{\rm form}$, but in a weakly coupled theory,
it is formed only with a probability $\alpha_s$. Therefore,
at the scale $k_{\rm form}$, there are $O(\alpha_s)$ quanta per logarithmic phase space
due to medium-induced parton branching. 
At scales below $k_{\rm form}$ (denoted as the LPM-region in Fig.~\ref{fig1}) the formation time is faster $t_{\rm form}< t$, 
and as a result of this the medium-induced splittings become more and more abundant as one 
moves from the scale $k_{\rm form}$ to an increasingly softer scale $p < k_{\rm form}$.
There is an $O(\alpha_s)$
probability of emitting a splittee at the scale $p$ at every $t_{\rm form}(p)$ and thus
the probability of finding a splittee with a momentum $p$ with $p<k_{\rm form}$ is parametrically%
\footnote{
At leading order, for $p\ll Q$, the prefactor can be extracted, \emph{e.g.}, from the 
splitting function of \cite{Baier:2000sb,Arnold:2008zu}, and in numerical form the spectrum reads
\begin{equation}
\frac{dP_{\rm find}(t)}{d\log\,p}\approx \frac{1}{\pi} C_A \alpha_s p^{-1/2} \hat{q}^{1/2}(p)t.
\end{equation}
}
\begin{equation}
\frac{dP_{\rm find}(t)}{d\log \, p} \sim \alpha_s \, t/t_{\rm form}(p) \sim \alpha_s \, \hat{q}^{1/2} p^{-1/2} \, t \,.
\label{eq2}
\end{equation}

While $t_{\rm form}(p)$ determines the minimal duration for a quantum to 
be created with probability $O(\alpha_s)$, the parametrically longer time 
\begin{equation}
t_{\rm split}(p) \sim t_{\rm form}(p)/\alpha_s
\end{equation} 
is needed to create this quantum with probability $\sim 1$. 
At fixed time $t$, the quanta that are created thus with $O(1)$ probability live 
at the scale $p \sim k_{\rm split}$
\begin{equation} 
 k_{\rm split}(t) \sim
\alpha_s^2 k_{\rm form}(t) \sim \alpha_s^2\, \hat{q}\, t^2\, ,
\end{equation}
which marks the end of the LPM-region in Fig.~\ref{fig1}.

%%%%%%%%%%%%%%%%%%%%%%%%%%%%%%%
\subsection{Medium Cascade}

Not all quanta that are created will stay where they were created. Those modes that have time to 
lose a significant fraction of their energy will cascade to a significantly lower scale $p$. 
For LPM-type radiation, the splitting that degrades energy the most is the hardest splitting and the hardest
splitting available is the quasi-democratic one.\footnote{Splittings where both splittees carry $O(1)$ of the parent momentum will be referred to as 'quasi-democratic' in the following.}

The timescale for a quasi-democratic splitting is of the 
same order of magnitude as the splitting time $t_{\rm split}(p)$ at the same scale. 
This is so because, in a non-abelian theory, the parametric emission time for LPM-type radiation is
independent of the momentum of the parent and is set by the momentum of the softer splittee, which for a quasi-democratic splittings is parametrically of the same order as the parent momentum. 

A quasi-democratic splitting
moves the energy deposited at scale $p$ to a lower scale of $p/2$. Parametrically
the lower scale, $p/2$, is of the same order as the original $p$, and therefore even a
democratic splitting is a local process in the logarithmic $p$-space. 
However, if a given mode has had time to undergo a quasi-democratic splitting,
then a successive quasi-democratic splitting of its daughters will take place on a shorter
time scale. Therefore, the scales which have had time to undergo a quasi-democratic
splitting with $O(1)$ probability can cascade all the way to the scale $T$  of the thermal bath within the same epoch $t$ within which the first democratic splitting occurred \cite{Blaizot:2013hx} (in the context of thermalization, the analogous process was discussed in~\cite{Baier:2000sb,Kurkela:2011ti,Kurkela:2011ub}).

To quantify these considerations, we introduce the time $t_{\rm res}(p)$
that a quantum resides at scale $p$ before splitting further. For $p > k_{\rm split}$,
this residence time must equal the lifetime $t$ of the system, and for lower scales $p$
it must scale like the formation time. We therefore have
\begin{equation}
t_{\rm res}(p) \sim
\left\{
\begin{array}{ll}
t\, ,\qquad &\hbox{for\, $k_{\rm form}> p> k_{\rm split}$, }\\
\sqrt{\frac{p}{k_{\rm split}}}t, \quad &\hbox{for\, $p< k_{\rm split}$}\, .
\end{array}
\right.
\end{equation}

The energy $\epsilon$ in this cascade is dominated by the hardest scale that can 
cascade, $\epsilon = n(k_{\rm split}) k_{\rm split} \approx k_{\rm split}$. 
This energy will move through all scales $p$
down to $T$ via quasi-democratic splittings, and since quanta spend a shorter time 
$t_{\rm res}(p)$ at lower momentum scale, only the fraction $t_{\rm res}(p)/t$ of 
quanta that arrived within the residence time will not have left already and will 
therefore contribute to the energy at the scale $p < k_{\rm split}(p)$,
\begin{equation}
\frac{d\epsilon}{d\log p} =  p\, \frac{dn}{d\log p} \sim k_{\rm split} \frac{t_{\rm res}(p)}{t} \sim \alpha_s \sqrt{\hat{q}\, p} t\, .
\label{eq8}
\end{equation}
By coincidence, the energy distribution (\ref{eq8}) in the region of the medium cascade
matches the distribution in the LPM region since 
$k_{\rm split} \frac{t_{\rm res}(p)}{t} \sim p\, t/t_{\rm split}(p)$. Accordingly, also
the number of quanta per $\log p$ shows the same $1/\sqrt{p}$-dependence as in the LPM region, 
\begin{equation}
\frac{dn}{d\log p} = \frac{1}{p} \frac{d\epsilon}{d\log p} \sim \alpha_s \sqrt{\qhat} t/\sqrt{p}\, .
\end{equation}
This similarity is particular to the $\sqrt{p}$ power law. We emphasize that despite these
similarities, the physics in the region of the medium cascade and in the LPM region are 
somewhat different.

\subsection{Medium cascade of the vacuum quanta}
The quanta produced in the DGLAP region will also undergo medium
interactions that will make the quanta radiate and split. The 
distribution of radiation from the vacuum quanta is the same as from
any other mode. Therefore the distribution of the daughters originating
from the medium quanta above $k_{\rm split}$ is
\begin{equation}
\frac{d P_{\rm find}}{d \log\, p \, d \log \, \theta} \sim \alpha_s \frac{t}{t_{\rm split}(p)}\, ,
\label{eq11}
\end{equation}
where $\alpha_s$ is just the number of vacuum quanta per double logarithmic unit
of phase space.
Again, the ratio $t/t_{\rm split}(p)$ is smaller than 1 for modes above $k_{\rm split}$,
and therefore the number of daughters is smaller than the number of vacuum splitted quanta. 

Below $k_{\rm split}$, however, both $t/t_{\rm split}(p)>1$ and $t/t_{\rm res}(p)>1$, leading to 
a cascade that is similar to the medium cascade discussed above. With the same arguments 
the number of quanta becomes
\begin{equation}
\frac{d n}{d \log\, p \, d \log \, \theta} \sim \alpha_s \frac{t}{t_{\rm split}(p)} \quad \textrm{for } p<k_{\rm split}(p). 
\end{equation}

%%%%%%%%%%%%%%%%%%%%%%%%%%%%%%%%%%%%%%%%%%%%%%
\subsection{Absence of Bethe-Heitler region}
We have argued that from $k_{\rm form}$ down to the momentum scale $p$, particle 
radiation is LPM suppressed. Given that formation times decrease with decreasing $p$,
one may wonder why there is not a softer momentum scale at which individual quanta
in the heat bath are resolved and an un-suppressed Bethe-Heitler radiation pattern 
$n(p) \propto 1/p$ off independent scattering centers results. 
To discuss this possibility, we need to introduce microscopic characteristics of the
perturbative medium that we have in mind. The small angle scattering time in this
medium will be $\tau_{\rm scatt} \sim 1/ (\alpha_s T)$, and $\hat{q} \sim \alpha_s^2\, T^3$ \cite{Braaten:1989mz,Aurenche:2002pd,Arnold:2008vd}.
The cross-over to a Bethe-Heitler region  is then expected to take place at the scale $k_{\rm BH}$ determined by
\begin{equation}
 	t_{\rm form}(k_{\rm BH}) \sim \tau_{\rm scatt}\, ,
\end{equation}
which implies
\begin{equation}
k_{\rm BH} \sim T\, .
\end{equation}
Therefore, only the infrared tail of the medium cascade has an $O(1)$
Bethe-Heitler correction whereas the LPM-suppressed splitting gives a good
description of the radiation at all higher scales $p>k_{\rm BH} \sim T$. 

%%%%%%%%%%%%%%%%%%%%%%%%%%%%%%%%%%%%%
\subsection{Angular distribution}
\label{sec:F}
We discuss now the angular distribution of quanta on the different momentum scales $p$.
There are two mechanisms. 
First, multiple soft scattering gives rise to transverse Brownian motion and 
determines the distribution at small angles. Second, rare large angle scattering leads to deviations
from Brownian motion that were first described by Moli\`ere for QED \cite{M47}. 
This Moli\`ere scattering will put quanta at large angular scales that cannot be 
reached by Brownian motion in time $t$.

\subsubsection{Angular distribution at small $\theta$}
In general, transverse Brownian motion moves quanta by an angle 
\begin{equation}
\theta_{\rm BR}^2 \sim \frac{k_\perp^2}{p^2} \sim \frac{\qhat \, t_{\rm res}(p)}{p^2} \, ,
\end{equation}
that is set by the time $t_{\rm res}(p)$ that the quantum resides and thus broadens at scale $p$. 
Quanta with  $p>k_{\rm split}$ will have spent a time of order of the duration of the 
jet evolution at $p$, that means, $t_{\rm res}(p)\sim t$. Therefore, these quanta reach a typical 
angle 
\begin{equation}
 	\theta_{\rm BR}(p) \sim \frac{\sqrt{\hat q\, t}}{p}\qquad \hbox{for $ k_{\rm form} > p>k_{\rm split}$.}
\end{equation}
This is the limiting angle up to which the LPM region extends in Fig.~\ref{fig1}.
This angle reaches from 
\begin{equation}
	\theta_{\rm BR}(k_{\rm form}) \sim \frac{1}{\sqrt{k_{\rm form} t}}
\end{equation}
at the upper bound of the LPM region to a parametrically larger angle 
\begin{equation}
	\theta_{\rm BR}(k_{\rm split}) \sim \frac{1}{\sqrt{\alpha_s^2\, k_{\rm split} t}}
	\sim \frac{1}{\alpha_s^2\sqrt{k_{\rm form} t}}
\end{equation}
at the lower bound. For $p < k_{\rm split}$, where resplitting happens, 
the time $t_{\rm res}(p)$  a quantum stays at the scale $p$ before leaving
by undergoing a quasi-democratic splitting is shorter than the lifetime of the
system, $t_{\rm res}(p)<t$, and this shortens the typical angle reached by 
transverse Brownian motion in the region of the medium cascade, 
\begin{equation}
\theta_{\rm BR}^2(p) \sim \frac{\hat{q}\, t_{\rm res}(p)}{p^2} 
\sim \frac{\hat q\, t}{p^{3/2} k_{\rm split}^{1/2} } \qquad \hbox{for $p < k_{\rm split}$}\, .
\label{eq16}
\end{equation}
It is remarkable that in the region of the medium cascade this angle does not
change with evolution time $t$, although many other scales in Fig.~\ref{fig1} do.
In fact, if one adopts the perturbative estimate $\hat{q} \sim \alpha_s^2\, T^3$, one 
finds that this angle is set by the simple ratio of momentum scales,
\begin{equation}
	\theta_{\rm BR}(p) \sim \left(\frac{T}{p}\right)^{3/4}\, .
	\label{thetaBR}
\end{equation}
This way of rewriting equation (\ref{eq16}) makes it also clear that $p \sim T$ is 
the largest scale that isotropizes.

Since Brownian motion leads to a Gaussian distribution in $d^2\theta$, which is 
parametrically flat for $\theta < \theta_{\rm BR}$,
the energy contained in this region in a double logarithmic unit of phase space is  
\begin{eqnarray}
	\frac{d\epsilon}{d\log p\, d\log\theta} % &=& 
	% p \frac{dP_{\rm create}}{d\log p\, d\log\theta}
%	\nonumber \\
	&\sim& p\, \frac{t}{t_{\rm split}(p)}\, \frac{\theta^2}{\theta_{\rm BR}^2}\,. 
	%	{\rm min}\left[1;t_{\rm res}(p)/t \right]\, .
		\label{eq18}
\end{eqnarray}

Before concluding this subsection, we point out as an aside 
that there is a logarithmic enhancement at 
small $\theta$. The reason is that the particles created later have had a shorter time to
broaden and this effect accumulates quanta in the collinear region. To be specific, consider 
the contribution of some last small time interval $\alpha_s^b t$ to the radiation.
It is
\begin{equation}
\frac{dP_{\rm create}}{d\log(p)} \sim \alpha_s^b \frac{t}{t_{\rm split}(p)}
\end{equation}
and these radiated quanta appear under the angle 
\begin{equation}
\theta_b^2 \sim \qhat \alpha_s^b t/p^2\, .
\end{equation}
Therefore in the differential probability distribution, the $\alpha_s^b$'s 
cancel, and all logarithmic times make an equally large contribution to all
logarithmic bins they can reach. Therefore, one expects to see a
logarithmic enhancement of equation (\ref{eq18}) in the collinear region.

\subsubsection{Angular distribution in the Moli\`ere scattering region}
So far, our estimates did not rely on assumptions about the microscropic structure of the
medium. If hard scatterings resolve partonic constituents in the medium, then these
rare occurences can move a quanta to angles $\theta > \theta_{\rm BR}$ \cite{D'Eramo:2012jh}. At 
scale $p\in [T,k_{\rm form}]$, there are $t/t_{\rm split}(p)$ quanta per $\log p$ and each of them 
will undergo a hard momentum transfer $q_\perp = p\, \theta$ during their stay at the 
momentum scale $p$ with probability \cite{Braaten:1989mz,Aurenche:2002pd,Arnold:2008vd}
\begin{equation}
\frac{dP_{\rm kick}}{d\log\theta} \sim \alpha_s^2 \frac{n_T}{p^2\, \theta^2}\, t_{\rm res}(p)
\end{equation} 
where $n_T\sim T^3$ is the number density of resolvable scattering centers. Multiplying these numbers, we get the 
probability to find a quantum at large angles%
\footnote{
Another possibility to create a quantum at large angles would be directly
through a hard scattering. The probability for this would then be $P_{\rm find}\sim \alpha_s \times P_{\rm kick}$.  However, whenever a quantum can be formed, $t/t_{\rm split}>\alpha_s$, and this process gives only a subleading correction.
}
\begin{eqnarray}
\frac{d P_{\rm find} }{d \log\,  p\,  d\log\theta} \sim
\left\{ \begin{array}{ll}
 \alpha_s^3 n_T t^2 \hat{q}^{1/2} /\left( p^{5/4}\, \theta\right)^2 & \textrm{ for } p>k_{\rm split}, \\
 \alpha_s^2 n_T t/(p \, \theta)^2 & \textrm{ for } p<k_{\rm split}.
 \end{array}
 \right.
 \label{eq22}
\end{eqnarray}
We find that along lines $p \propto \theta^{-4/5}$ for $p> k_{\rm split}$ (or $p \propto \theta^{-1}$ for $p< k_{\rm split}$), there is a fixed number of quanta per double logarithmic phase space. 
Since this is a power-law dependence in $\theta$ while
Brownian motion dies out exponentially above $\theta_{\rm BR}$, rare hard Moli\`ere scattering,
if allowed for by the microscopic structure of the medium, will dominate the distribution of medium-induced
quanta above $\theta_{\rm BR}$. 

Comparing the probability distribution (\ref{eq22}) for Moli\`ere scattered quanta at large angle 
$\theta > \theta_{\rm BR}(p)$ to that of vacuum quanta (eq.~(\ref{eq1}) for $p > k_{\rm split}$)
and to that of vacuum quanta undergoing the medium cascade (eq.~(\ref{eq11}) for $p < k_{\rm split}$), 
we find that outside a relatively narrow region of logarithmic 
phase space, the DGLAP vacuum radiation 
(thin dash-dotted line in Fig.~\ref{fig1}), or the cascade of the vacuum quanta (thick dash-dotted line in 
Fig.~\ref{fig1}), will overshadow the contribution from Moli\`ere scattering. 
The quanta sensitive to the microscopic structure give a dominant contribution to the energy only for
$\theta < \theta_{\rm Mol}$ with
\begin{equation}
\theta_{\rm Mol} \sim 
\left\{
\begin{array}{ll}
\alpha_s \, t \, n_{T}^{1/2} \, \hat{q}^{1/4}p^{-5/4} & \textrm{ for } p>k_{\rm split}\, ,   \\
n_T^{1/2} \hat{q}^{-1/4} p^{-3/4}& \textrm{ for } p<k_{\rm split}\, .
\end{array}
\right.
\end{equation}
Here, the expression for $p<k_{\rm split}$ can be written in a form resembling equation (\ref{thetaBR})
\begin{equation}
\theta_{\rm Mol} \sim \frac{1}{\alpha_s^{1/2}}\left( \frac{T}{p}\right)^{3/4} \textrm{ for } p<k_{\rm split},
\end{equation}
As already observed for $\theta_{\rm BR}$, also the angular limit $\theta_{\rm Mol}$ for 
Moli\`ere scattering is independent of time within the region of the medium cascade.

%%%%%%%%%%%%%%%%%%%%%%%%%%%%%%%%%%%%%%%%%%%%%%%%
\subsection{Extending Fig.~\ref{fig1} to $t > t_{\rm form}(Q)$}
So far in our consideration, we have made the assumption that $Q$ is the hardest scale in the system. The scales $k_{\rm form}$ and $k_{\rm split}$, however, grow fast $\propto t^2$, and eventually they will reach the scale $Q$. In the following we discuss how the dynamics changes after $k_{\rm form}$ and $k_{\rm split}$ reach $Q$.

When $k_{\rm form}\sim Q$ (or equivalently $t>t_{\rm form}(Q)$) the quasi-democratic splitting of the original jet becomes allowed, but it happens only with probability $\alpha_s$; successive quasi-democratic splitting are also suppressed by further powers of $\alpha_s$. Therefore, for a typical jet, no qualitative change happens at $t_{\rm form}(Q)$. What changes, however, is the average rate at which the leading parton is losing energy $d\langle \epsilon_Q \rangle/dt$; here the brackets are to be understood as an average over an ensemble of independent jets. The energy lost by the leading parton is dominated by the hardest splitting possible, which before $t_{\rm form}(Q)$ is $k_{\rm form}$ but stays at $Q$ after $t_{\rm form}(Q)$. The average rate for losing energy depends on the probability of the hardest emission $t/t_{\rm split}$, and it depends on the energy lost in the event of a hard emission taking place. For $t<t_{\rm form}(Q)$ the average rate reads~\cite{BDMPSZ}
\begin{equation} 
d\langle \epsilon_{Q }\rangle/d t \sim (t/t_{\rm split}(k_{\rm form})) k_{\rm form}/t 
%\sim \alpha_s k_{\rm form}/t 
\sim \alpha_s \hat{q} t\, ,
\label{eq25}
\end{equation} 
whereas the rate saturates for $t>t_{\rm form}(Q)$
\begin{equation} 
d\langle \epsilon_{\rm Q}\rangle/d t \sim Q t/t_{\rm split}(Q) \sim \alpha_s Q^{1/2}\hat{q}^{1/2}.
\label{eq26}
\end{equation} 

When $k_{\rm split}\sim Q$ --- corresponding to the jet stopping time of $t_{\rm split}(Q)\sim (Q/T)^{1/2}/\alpha_s^2 T$ \cite{BDMPSZ,Arnold:2009ik}  ---the probability for a democratic splitting of the 
jet becomes $O(1)$, and the successive democratic splittings will happen in 
a time scale faster than the time it took for the original splitting. As argued, \emph{e.g.}, in 
Refs.~\cite{Baier:2000sb,Arnold:2009ik,Kurkela:2011ti,Kurkela:2011ub,Kurkela:2014tea,Blaizot:2013hx}, the jet will therefore become quenched and lose all of its energy to the thermal bath in a timescale it takes to 
undergo the first quasi-democratic splitting.

\section{Jet Quenching in a finite medium}
The picture Fig.~\ref{fig1} discussed in section~\ref{sec2} is a snapshot at fixed evolution time 
of the logarithmic phase space distribution of partonic fragments. 
Some of the phase space boundaries will evolve in time as indicated in the figure, while others
will stay put. Therefore, this picture informs us also about how the jet quenching process
occurs dynamically as a function of time. 

In the phenomenologically realized situation, the jet propagates over a finite path-length $L$ in 
QCD matter, and (for sufficiently short $L < t_{\rm form}(Q)$, see details below) 
Fig.~\ref{fig1} thus represents the distribution of partonic jet fragments at 
moment $t\sim L$ when the jet escapes the medium. In the following, we discuss in more
detail the energy distribution of the quenched parton shower at that moment, and the 
physics that modifies this distribution at later times and for longer path lengths.

\subsection{Vacuum radiation for $t > L$}
The typical virtuality of quanta at time $t\sim L$ is 
$O\left(\hat{q}\, t_{\rm res}(p)\right)$. The dominant mechanism that further degrades this
virtuality in the subsequent vacuum evolution is soft and collinear splitting. On the one hand,
this mechanism does not move appreciably the emitter in $\log\theta$ and $\log p$, on the
other hand, it puts $O(\alpha_s)$ quanta in the double logarithmic phase space below any 
emitter. Therefore, late time fragmentation affects the distributions of Fig.~\ref{fig1} only 
in those regions of phase space in which there are less quanta than $O(\alpha_s)$ times the 
number of quanta at higher momentum scales in the corresponding angular scale 
at time $t \sim L$. As we explain now, this is only the case in the collinear region of sufficiently 
small $\log \theta$, where 
momentum broadening and smallness of angular phase space have resulted in a small density
of medium-induced quanta, and in a sufficiently soft region at large angle where there are
sufficiently many medium-induced quanta at higher momentum scale that can split further at late times.

\subsubsection{Radiation of vacuum quanta at $t>L$}
\label{sec3a1}
At small angular scales $\theta  < \theta_{\rm BR}(k_{\rm split })\sim (\alpha_s^2 k_{\rm split}t)^{-1/2}$, there are less than 1 medium
induced quanta integrated over $p$ and therefore late vacuum fragmentation is dominated by 
the original single parton at the scale $Q$. This becomes dominant over the medium-induced
distribution when there are less than $O(\alpha_s)$ medium induced quanta per double logarithmic
unit of phase space. This small-angle region $\theta(p) < \theta_{\alpha}(p)$, in which
vacuum radiation dominates is obtained by requiring that (\ref{eq18}) is larger than $\alpha_s\, p$, 
\begin{equation}
	\theta_{\alpha}(p) \sim \sqrt{\alpha_s}\, \theta_{\rm BR}(p)\, \sqrt{\frac{t_{\rm split}(p)}{t}}\, .
\end{equation}
This region is delineated by the thin dashed line between $k_{\rm split}$ and $k_{\rm form}$ in Fig.~\ref{fig1}. We note that the contribution of this
late time vacuum splitting to jet observables will be that of a vacuum jet, but one of degraded energy. 
For an ensemble average, the reduced energy of this vacuum jet contribution is given by equations (\ref{eq25}) and (\ref{eq26}), 
\begin{equation}
	\langle Q'\rangle = Q - \langle \epsilon_{Q }\rangle \sim Q -
	\left\{
	\begin{array}{ll}
	\alpha_s k_{\rm form}&\textrm{ for }t<t_{\rm form}(Q)\\
	\alpha_s Q^{1/2}\hat{q}^{1/2}&\textrm{ for }t>t_{\rm form}(Q).
	\end{array}
	\right.
\end{equation}
In principle, late time splitting
of vacuum quanta contributes also to $\theta > \theta_{\rm BR}(k_{\rm split})$. In this region of the medium-cascade, 
however, there is also a contribution from the late time vacuum splitting of medium-induced quanta
to which we shall turn next. Since this latter contribution dominates, we do not continue in Fig.~\ref{fig1}
the line of $\theta_\alpha(p)$ into the region $\theta > \theta_{\rm BR}(k_{\rm split})$ corresponding to $p < k_{\rm split}$.

%%%%%%%%%%%%%%%%%%%%%%%%%%%%%%%%%%%
\subsubsection{Radiation of medium-induced quanta at $t>L$}
\label{sec3a2}
Most of the medium-induced quanta reside on the angular scale $\theta_{\rm BR}(p)$. 
Correspondingly, for a given angular scale $\theta > \theta_{\rm BR}(k_{\rm split})$, the number of medium-induced quanta is dominated by the momentum scale
\begin{equation}
p_{\rm BR}(\theta) \sim \left( \frac{\theta_{\rm BR}}{\theta}\right)^{4/3}p\, . 
\end{equation}  
For these angular scales, the number of the medium-induced 
quanta along $\theta_{\rm BR}(p)$ exceeds $O(1)$ per double logarithmic phase 
space. Therefore, the late time vacuum splitting of these more than $O(1)$ medium-induced quanta 
dominates over the vacuum splitting of the $O(1)$ vacuum quantum described in section~\ref{sec3a1} 
above.

For a fixed $\theta$, the number of quanta at the scale $p_{\rm BR}$ reads
\begin{equation}
\frac{dn}{d\log p\,d\log \theta}\sim \frac{t}{t_{\rm split}(p)}  \frac{t_{\rm split}(p)}{t_{\rm split}(p_{\rm BR})}\sim \frac{t}{t_{\rm split}(p)} \left( \frac{\theta}{\theta_{\rm BR}}\right)^{2/3}.
\end{equation}
Then the number of emitted quanta by the medium induced fragments at the scale $p_{\rm BR}$ to all 
scales below is
\begin{equation}
\frac{d n_{\rm late\, vac}}{d \log p\, d \log \theta} \sim \alpha_s \frac{t}{t_{\rm split}(p)} \left( \frac{\theta}{\theta_{\rm BR}}\right)^{2/3}.
\end{equation}
This dominates over the medium-induced quanta for
\begin{equation}
\theta < \theta_{\alpha} \sim \alpha_s^{3/4}\theta_{\rm BR}\quad\textrm{ for } \theta > \theta_{\rm BR}(k_{\rm split})\, .
\end{equation}
This condition extends the
$\theta_\alpha$-line to larger angular scales $\theta_\alpha>\theta_{\rm BR}(k_{\rm split})$. We have
denoted this extension by a thicker dashed purple line in Fig.~\ref{fig1} to emphasize that there
are more than $O(\alpha_s)$ quanta along this line for $\theta_\alpha>\theta_{\rm BR}(k_{\rm split})$.

\subsection{Difference in energy degradation of leading hadrons and jets}
Since leading hadrons are the leading fragments of leading partons, the total energy 
$\alpha_s k_{\rm form}$ radiated away from $Q$ up to time $t$ due to medium effects
sets the scale for the suppression of leading hadrons.\footnote{As explained in Ref.\cite{Baier:2001yt}, 
the typical energy shift seen in leading hadron spectra can be significantly smaller than
$\alpha_s k_{\rm form}$ due to trigger bias effects.} In contrast, what is radiated away from the 
leading parton is not necessarily radiated outside the phase space within which the
jet is reconstructed. It is only the energy in scales $p < k_{\rm split}$ that has had
time to undergo the medium cascade and that therefore escapes for sure
to sufficiently large angles \cite{Baier:2000sb,Kurkela:2011ti,Blaizot:2013hx}. The energy missing from the reconstructed jet is thus given by
\begin{equation}
\epsilon_{\rm jet} \sim k_{\rm split} \sim \alpha_s^2 \hat{q}t.
\end{equation}
Fig.~\ref{fig1} therefore implies that for all in-medium path lengths $L < t_{\rm split}(Q)$, 
$d\langle \epsilon_{Q }\rangle/dt >  d \epsilon_{\rm jet }/d t$, and thus leading hadrons
are more suppressed than reconstructed jets. It is only at the time $t \sim t_{\rm split}(Q)$
when $k_{\rm split} = Q$ that $\langle \epsilon_{Q }\rangle \sim  \epsilon_{\rm jet}$.

%
%%%%%%%%%%%%%%%%%%%%%%%%%%%%%%%%%%%%%%%%%%
\begin{figure}[t]
\begin{center}
\vspace{.5cm}
\includegraphics[width=0.377\textwidth]{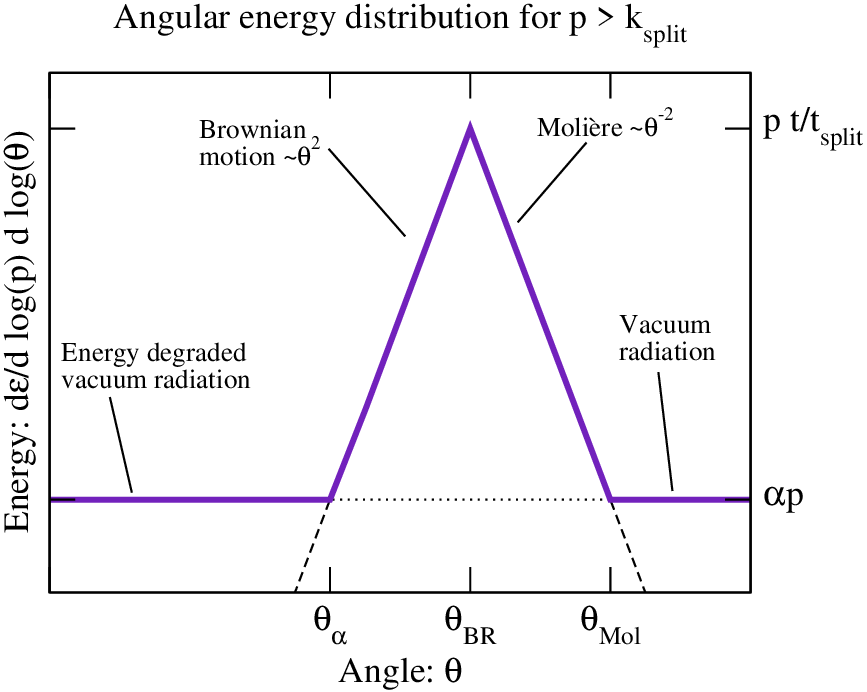}
\caption{
Double logarithmic sketch of the distribution of energy as a function of angle for a fixed $p>k_{\rm split}$. The large angular
scales $\theta>\theta_{\rm Mol}$ are dominated by DGLAP vacuum radiation from the leading parton at the scale
$Q$. At small angular scales $\theta<\theta_\alpha$ the energy density is dominated by vacuum radiation of the 
leading parton after it has degraded its energy propagating through the medium. Medium induced radiation is
centered around the angular scale $\theta_{\rm BR}$: for $\theta_\alpha < \theta< \theta_{\rm BR}$ the angular 
spectrum is that of Brownian motion whereas for $\theta_{\rm BR} < \theta< \theta_{\rm Mol}$ the spectrum arises from 
Moli\`ere scattering. The medium induced contributions (dashed lines) grow as a function of evolution time with respect to the vacuum one (dotted line).
\label{fig2}
}
\end{center}
\end{figure}
\begin{figure}[t]
\begin{center}
\vspace{.5cm}
\includegraphics[width=0.4\textwidth]{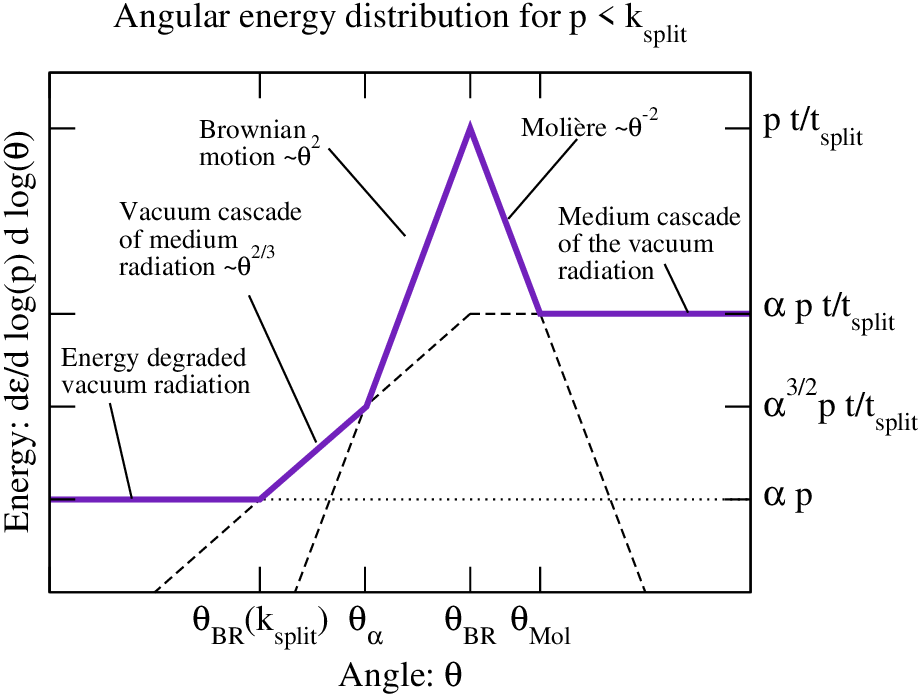}
\caption{
Double logarithmic sketch of the distribution of energy as a function of angle for a fixed $p< k_{\rm split}$. 
At large angles $\theta>\theta_{\rm Mol}$, in-medium fragmentation of DGLAP radiation dominates the energy density. In close analogy to the situation in the LPM region depicted in Fig.~\ref{fig2}, there is a small angle
region dominated by (energy degraded) late DGLAP vacuum radiation, and there is the physics of 
medium-induced Brownian motion and large angle Moli\`ere scattering that dominates to the left and to the right of $\theta_{\rm BR}$, respectively. In addition, there is a window $\theta_{\rm BR}(k_{\rm split}) < \theta < \theta_\alpha$ in which the late time vacuum DGLAP radiation originating from the medium-induced splittees dominates the spectrum. This window closes for $p > \alpha_s^{3/2}\, k_{\rm split}$ as described in the text. 
The medium induced contributions (dashed lines) grow as a function of evolution time with respect to the vacuum radiation (dotted line).
\label{fig3}
}
\end{center}
\end{figure}

%%%%%%%%%%%%%%%%%%%%%%%%%%%
%
\subsection{Angular jet energy distribution}
The parametric estimates for the angular distribution of jet energy obtained in this note
can be written in a compact form for $p<k_{\rm form}$
\begin{eqnarray}
&&\frac{d \epsilon}{d \log \, p \, d \log \, \theta} \sim
p\,\frac{t}{t_{\rm split}(p)}
\nonumber \\
&& \quad \times
 \left\{
\begin{array}{ll}
\alpha_s (t_{\rm split}(p)/t) & \textrm{ for } \theta < \theta_{\rm BR}(k_{\rm split})\, , \\ 
\alpha_s (\theta/\theta_{\rm BR})^{2/3} & 
\textrm{ for } \theta_{\rm BR}(k_{\rm split})<\theta< \theta_\alpha\, ,\\
 (\theta^2/\theta_{\rm BR}^2) & \textrm{ for } \theta_\alpha < \theta < \theta_{\rm BR}\, ,\\
 (\theta_{\rm BR}^2/\theta^2) & \textrm{ for }  \theta_{\rm BR} < \theta  <\theta_{\rm Mol}\, ,\\
 \alpha_s t_{\rm split}(p)/t_{\rm res}(p) & \textrm{ for }  \theta  >\theta_{\rm Mol}\, .
\end{array}
\right.
\label{eq29}
\end{eqnarray}
The expression 
(\ref{eq29}) is valid for both, the LPM region, $k_{\rm split}<p<k_{\rm form}$, and 
the region of the medium cascade, $p<k_{\rm split}$. The second of the five conditions listed
in (\ref{eq29}) is realized only when $ \theta_{\rm BR}(k_{\rm split}) <  \theta_\alpha$,
which corresponds to $p<\alpha_s^{3/2} k_{\rm split}$. For $\theta_\alpha < \theta_{\rm BR}(k_{\rm split})$,
the region dominated by late energy degraded vacuum radiation (first line in eq.(\ref{eq29})) extends up
to $\theta_\alpha$ and connects directly to the region dominated by Brownian motion of medium quanta.

The angular jet energy distribution (\ref{eq29}) is 
depicted in Figs.~\ref{fig2} (for the LPM-region) and~\ref{fig3} (for the medium cascade region). 
Figs.~\ref{fig2} and~\ref{fig3} illustrate that indeed, late time vacuum radiation of $O(\alpha_s)$ quanta
dominates at angles $\theta < \theta_\alpha(p)$. It is only in the range
$\theta_\alpha(p) < \theta < \theta_{\rm Mol}$ that purely medium-induced contributions
dominate. In the regions $\theta_{\rm BR}(k_{\rm split}) < \theta_{\alpha}$ and $\theta>\theta_{\rm Mol}$ 
the energy is dominated neither by pure vacuum radiation nor by medium-induced radiation, but
is a result of an interplay of both types of radiation. 
While one may naively have thought that the energy from large angle
Moli\`ere scattering appears at much larger angles than that from small angle
Brownian motion, Figs.~\ref{fig2} and~\ref{fig3} demonstrate that the dominant energy from both
contributions pile up at the same angular scale $\theta_{\rm BR}$. It is then only
the shape of the angular dependence between $\theta_{\rm BR}$ and $\theta_{\rm Mol}$
that may give access to microscopic details of jet-medium 
interaction~\cite{D'Eramo:2012jh}. 

We further emphasize  that $\theta_{\rm BR}, \theta_{\alpha}$, and $\theta_{\rm Mol}$ are independent of evolution time below $k_{\rm split}$, and that a characteristic medium-induced enhancement is 
seen at this angular scale. This angular scale is set directly by the temperature, while
the only medium-dependent information entering the size of the peak is $\hat q$. 
If this structure would be experimentally accessible, it would thus give direct
access to the temperature dependence of $\hat q$. More generally, Figs.~\ref{fig2} and~\ref{fig3}
illustrate that the energy per unit of double logarithmic phase space will 
peak for all momentum scales $p$ on the characteristic scale $\theta_{\rm BR}(p)$ that is a 
medium-induced scale. This is a robust expectation for perturbative mechanisms of jet quenching.
These parametric considerations may provide
a motivation to characterize experimental data on the angular jet energy distribution
in $\log\theta$ and to search for such an enhancement. 

\section{Conclusions}
The main deliverables of this paper are Figs.~\ref{fig1}, ~\ref{fig2} and ~\ref{fig3} which provide 
a unified view of the physics underlying jet quenching. As discussed, this view is consistent
with many results on jet quenching in the (parametrically recent) literature.  
We note that all the physics phenomena invoked in our discussion are 
at least in principle implemented in some of the documented jet quenching models.
The contribution of our note is not so much to point to novel physics effects, but to provide a
map of the phase space regions in which specific known physics is expected
to dominate. For instance, the characterization of the angular distribution of jet energy 
in Figs.~\ref{fig2} and ~\ref{fig3} points to the interest in searching in $\log \theta$ plots 
(within Monte Carlo studies and within data) for characteristic medium-induced enhancements 
at scale $\theta_{\rm BR}$ that may inform us not only about the temperature and small-angle 
scattering properties ($\hat q$) of the medium, but also (if one can identify the region of Moli\`ere
scattering) about its quasi-particle content~\cite{D'Eramo:2012jh}.
We hope that in this and in other ways, the simple Figs.~\ref{fig1}, ~\ref{fig2} and ~\ref{fig3}
will be of use in the further discussion of jet quenching phenomena.

\acknowledgements
We thank J.P. Blaizot, J. Ghilglieri, E. Iancu, Y. Mehtar-Tani, A.H. Mueller and K. Zapp for 
discussions at various stages of this work.

\end{document}